\documentclass{article}
\usepackage[T1]{fontenc}
\usepackage{spconf,amsmath,graphicx,hyperref}
\usepackage{amsfonts}
\usepackage{arydshln}
\usepackage{booktabs}
\usepackage{xcolor}
\newlength{\tableentrywidth}
\newcommand{\tabbest}[1]{\begingroup
  \settowidth{\tableentrywidth}{$#1$}%
  \resizebox{\tableentrywidth}{\height}{$\mathbf{#1}$}\endgroup}

\title{MIDI-REPA: Improving Text-to-Music Generation with Symbolic Representation Alignment}
\name{Song-Ze Yu$^{1}$, Fang-Duo Tsai$^{1,2}$, Yi-Hsuan Yang$^{1}$}
\address{$^1$National Taiwan University $^2$University of California, San Diego}

\begin{document}
\ninept
\maketitle
\begin{abstract}
State-of-the-art text-to-music models increasingly rely on scaling model capacity, data, and compute, while still learning musical structure directly from dense audio. Yet melody, rhythm, harmony, and structure can be represented compactly in symbolic form. We introduce \textbf{MIDI-REPA}, a representation-alignment framework that derives MIDI supervision from automatic music transcription (AMT) to align intermediate representations of a flow-matching DiT. Building on the 122M-parameter FluxAudio-S baseline from the 2026 ICME Academic Text-To-Music Challenge, we investigate \textit{what}, \textit{where}, and \textit{when} to align. Aligning a MuScriptor-derived piano roll with an early DiT layer throughout the full flow trajectory performs the best. Without additional training data or techniques, MIDI-REPA remains competitive on the ATTM objective metrics while remaining trainable on a single RTX 3090 in about 31 hours. Our results highlight symbolic supervision as a simple and inexpensive learning signal for improving text-to-music generation.  We release our transcription pipeline, JamendoMIDI, training code, and audio samples at \url{https://midi-repa.vaclis.net}.

\end{abstract}

\section{Introduction}
\par\vspace{-0.3em}
\label{sec:intro}
Text-to-music generation has made rapid progress in audio quality and musical coherence, that is already difficult for listeners to distinguish from human-made music\cite{deezer2025aisurvey}. Yet such progress largely relies on massive scaling of model capacity, training data, and compute \cite{gong2026ace,evans2026stable,yuan2026yue}, making state-of-the-art developments beyond the reach of most researchers and users.

From autoregressive audio tokens \cite{copet2023simple} to diffusion \cite{huang2023noise2music,evans2024fast} and flow-matching objectives \cite{prajwal2024musicflow}, models are asked to learn both \textit{musical content} (melody, rhythm, harmony, structure) and \textit{acoustic realization} (timbre, production) directly from the audio. However, musicians do not conceive and realize these two aspects of music in the same form. Musical content can be described compactly and structurally by a symbolic score with only a sparse sequence of discrete events, whereas acoustic realization is captured by a dense audio signal with millions of continuous-valued samples. Although audio partly encodes musical content, reconstructing or denoising dense audio representations forces models to learn sparse musical structure alongside far richer acoustic variation, which may underemphasize the former. Conditioning on musical content~\cite{wu2024music, tsai2025musecontrollite, tsai2026midi} helps, but musicality degrades without the condition. Large-scale pretraining may mitigate this, but does not show that audio is the best space for learning musical content.

Representation alignment approaches in computer vision have been shown to improve generative models training by guiding intermediate representations toward pretrained encoder features \cite{yu2024repa}. Recent text-to-music works extend this idea using pretrained audio encoders such as MERT, CLAP, and MuQ \cite{gong2025ace, cheng2026making}. However, these methods primaily align with learned audio representations, and to our knowledge, none directly align with structured musical content. This motivates a more fundamental research question beyond scale alone: \textit{Can text-to-music generation be improved by guiding its internal representations toward symbolic musical content?} 

To answer this question, we introduce \textbf{MIDI-REPA},  a representation-alignment framework that trains a text-to-music flow-matching model from scratch while aligning intermediate representations with symbolic music representations derived from existing automatic music transcription (AMT) systems. Following the 2026 ICME Academic Text-To-Music (ATTM) Challenge setup, we train on its instrumental MTG-Jamendo dataset and build on their provided Diffusion Transformer (DiT)-based FluxAudio-S baseline, without introducing additional training techniques \cite{hsieh2026academic}. We systematically investigate \textit{what}, \textit{where}, and \textit{when} to align: which AMT system and aligned representation should serve as the symbolic MIDI teacher, which DiT layer should be aligned, and over which timestep range along the flow trajectory the alignment should be applied. 

Our experiments show that symbolic musical content provides an effective learning signal for text-to-music generation. Our best configuration aligns a MuScriptor~\cite{rouard2026muscriptor} piano-roll with an early DiT layer (L2) throughout the full flow trajectory. This configuration remains competitive on the ATTM objective metrics, while keeping the backbone at only 122M parameters, using the same training data, and remaining trainable on a single RTX 3090 GPU in approximately 31 hours. As text-to-music models continue to scale, our findings highlight symbolic musical content should not be overlooked as a natural and direct source of supervision during training. 

\begin{figure*}[t]
  \centering
  \includegraphics[width=0.89\textwidth]{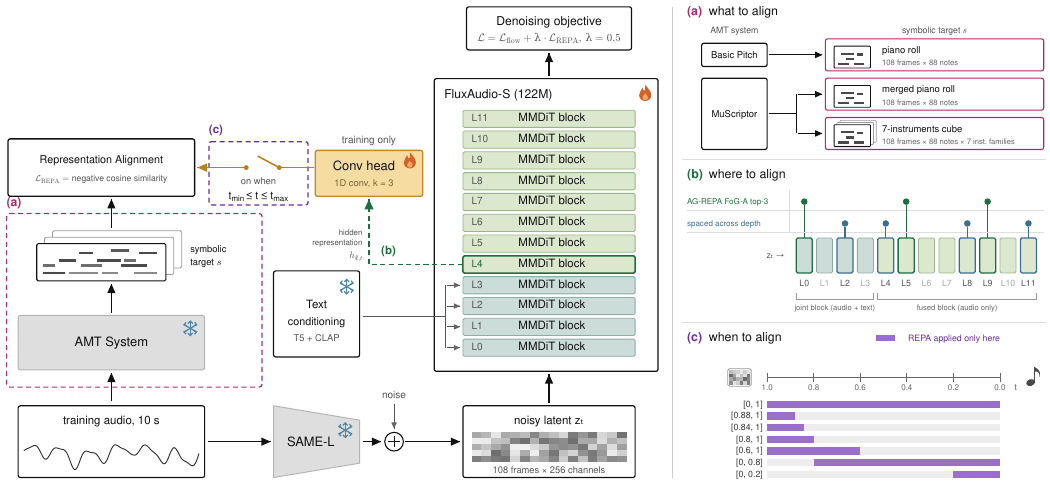}
  \caption{MIDI-REPA framework. Symbolic targets extracted by an AMT system supervise intermediate audio representations through a lightweight projection head and cosine-similarity alignment loss, jointly with the denoising objective. Our design study examines (a) \emph{what} symbolic target to align, (b) \emph{where} to apply alignment in the backbone, and (c) \emph{when} to activate it along the flow-matching trajectory.}
  \label{fig:framework}
  \par\vspace{-0.8em}
\end{figure*}

\par\vspace{-0.35em}
\section{Method}
\par\vspace{-0.35em}
\label{sec:framework}

\subsection{MIDI-REPA Framework}
\par\vspace{-0.35em}
As shown in figure~\ref{fig:framework}, for each training audio sample, we precompute a time-aligned symbolic target $\mathbf{s}$ offline using a pretrained automatic music transcription (AMT) system~\cite{rouard2026muscriptor,bittner2022lightweight} as detailed in Section \ref{subsub:target_extraction}. During flow-matching training, we extract the hidden representation $\mathbf{h}_{\ell,t}$ from an intermediate block $\ell$ of the
text-conditioned DiT at timestep $t$, and use a lightweight
single-layer temporal convolutional projector $g_{\phi}(\cdot)$ inspired by iREPA~\cite{singh2025matters} to
project $\mathbf{h}_{\ell,t}$ into the symbolic target $\mathbf{s}$ space:
$\hat{\mathbf{s}}_{\ell,t}=g_{\phi}(\mathbf{h}_{\ell,t})$. 

We retain the original conditional flow-matching objective $\mathcal{L}_{\mathrm{flow}}$ and introduce an auxiliary symbolic representation alignment objective $\mathcal{L}_{\mathrm{REPA}}$. Let $\mathcal{V}$ denote the set of frames with valid symbolic activity. For timesteps within the
alignment interval $\mathcal{T}$, the alignment objective is
the negative cosine similarity between the projected hidden
representation and the aligned symbolic target:
\begin{equation}
\mathcal{L}_{\mathrm{REPA}}
=
-\frac{1}{|\mathcal{V}|}
\sum_{i\in\mathcal{V}}
\operatorname{cos}
\left(
\hat{\mathbf{s}}_{\ell,t}^{(i)},
\mathbf{s}^{(i)}
\right).
\label{eq:repa}
\end{equation}

The overall training objective is
\begin{equation}
\mathcal{L}
=
\mathcal{L}_{\mathrm{flow}}
+
\lambda
\mathbb{I}[t\in\mathcal{T}]
\mathcal{L}_{\mathrm{REPA}},
\label{eq:objective}
\end{equation}
where $\lambda$ controls the strength of symbolic alignment.
The symbolic target $\mathbf{s}$, alignment block $\ell$, and
timestep interval $\mathcal{T}$ respectively determine
\textit{what}, \textit{where}, and \textit{when} to align.
Both the AMT system and the projector $g_{\phi}(\cdot)$ are used only during
training and introduce no additional inference-time computation.
We investigate these three design choices in Sec.~\ref{sec:design}.

\par\vspace{-0.1em}
\subsection{Experimental setup}
\subsubsection{Model architecture}
We instantiate the framework on FluxAudio-S, the official 122M-parameter baseline from the 2026 ICME Academic Text-to-Music (ATTM) Grand Challenge~\cite{hsieh2026academic, fei2024flux}. The backbone uses T5+CLAP conditioning and contains 12 MMDiT blocks: four joint audio/text blocks followed by eight fused blocks. We index these blocks as L0--L11.
We replace its original audio autoencoder with frozen SAME-L, a semantically regularized music autoencoder~\cite{parker2026same}, to reduce the risk that generation quality is bottlenecked by the audio latent representation. Ten seconds of 44.1-kHz stereo audio are encoded into latents $\mathbf{z}\in\mathbb{R}^{T\times C}$ with $T=108$ frames and $C=256$ channels.

\par\vspace{-0.3em}

\subsubsection{Symbolic target extraction}
\label{subsub:target_extraction}
We derive symbolic targets using two AMT systems: the instrument-agnostic
Basic Pitch~\cite{bittner2022lightweight} and the multi-instrument
MuScriptor~\cite{rouard2026muscriptor}. Their transcriptions are converted to binary time--pitch representations with $T=108$ frames and $P=88$ pitches (MIDI 21--108, A0--C8), matching the temporal resolution of the SAME-L latents. Basic Pitch yields a single piano roll, while
MuScriptor yields multi-instrument piano rolls that can be retained as an instrument cube or merged into a single piano roll.

In our preliminary experiments, MuScriptor produces unstable predictions
when conditioned on either no instrument tags or all instrument tags,
motivating the use of instrument conditioning. Since such metadata
are often unavailable in large-scale music datasets, we obtain the
conditioning set using multi-instrument recognition. We benchmark
PANNs, DyMN-L, MATPAC++, and Music Flamingo
~\cite{kong2020panns,schmid2024dynamic,quelennec2025matpac,ghosh2026music}
on MulttiPop~\cite{pruyne2026multtipop}. We observe that audio encoders
generally achieve higher recall, whereas the ALM provides higher precision.
Since MuScriptor hard-masks instrument tokens outside the conditioning, we prioritize recall and take the union of MATPAC++ and Music
Flamingo. To reduce confusion among closely related instruments, we map their predictions into $K=7$ instrument families: piano, guitar, bass, drums, strings, winds, and synthesizer. This yields a MuScriptor cube $\mathbf{s}\in\{0,1\}^{K\times T\times P}$; merging its instrument channels gives a merged piano roll $\mathbf{s}\in\{0,1\}^{T\times P}$. For the cube, each frame's target is flattened to $KP$ dimensions, and the projector output is sized accordingly.

\par\vspace{-0.3em}
\subsubsection{Training data and optimization}
\label{subsub:training_config}

We run our transcription pipeline on ATTM's 55,701 vocal-removed Jamendo tracks~\cite{hsieh2026academic} and discard 4,624 tracks with unusable outputs, yielding \textbf{JamendoMIDI}, 51,077 audio tracks paired with instrument-stem MIDI. For training, we further exclude 682 Song Describer Dataset (SDD) tracks~\cite{manco2023songdescriber} and 930 ATTM reference tracks from it contains, and reserve 444 tracks for loss monitoring, leaving 49,034 tracks ($\sim$3,161 hours).\footnote{SDD and ATTM share 13 tracks, all excluded from training.} We use SDD for design comparisons and ATTM for the final benchmark. For each training track, we randomly select one of the two challenge caption sources (Qwen and Flamingo) and keep the assignment fixed throughout training. All arms use the same data and initialization seed, AdamW, batch size 128, 300k updates, learning rate $10^{-4}$ with 1k warm-up steps and tenfold reductions at 240k and 270k, and $\lambda=0.5$. Noise levels follow $t=\operatorname{sigmoid}(u)$, $u\sim\mathcal N(0,1)$. Evaluation uses the final EMA ($\sigma_{\rm rel}=0.05$), 25 Euler steps, and classifier-free guidance 4.5.
 
\begin{table}[t]
\centering
\caption{What to align: Symbolic targets at L4 on SDD. Bold and underline indicate the best and second-best results; gray indicates performance worse than the baseline.}
\label{tab:teacher_comparison}
\small
\setlength{\tabcolsep}{3pt}
\renewcommand{\arraystretch}{1.02}
\begin{tabular*}{\columnwidth}{@{\extracolsep{\fill}}lcc@{}}
\toprule
Teacher & FAD $\downarrow$ & CLAP $\uparrow$ \\
\midrule
MuScriptor cube & $\underline{0.5117 \pm 0.0042}$ & $\tabbest{0.2665 \pm 0.0030}$ \\
MuScriptor roll & $\tabbest{0.5040 \pm 0.0030}$ & $\underline{0.2603 \pm 0.0026}$ \\
Basic Pitch roll & \textcolor{gray}{$0.5363 \pm 0.0036$} & $0.2396 \pm 0.0029$ \\
\midrule
Baseline & $0.5275 \pm 0.0021$ & $0.2220 \pm 0.0021$ \\
\bottomrule
\end{tabular*}
\end{table}
\begin{table}[t]
\centering
\caption{Baseline FoG-A over 512 training windows and five noise levels. Bold: top three contributions; these are not alignment-performance ranks.}
\label{tab:foga}
\small
\setlength{\tabcolsep}{3pt}
\begin{tabular*}{\columnwidth}{@{\extracolsep{\fill}}lrrlrr@{}}
\toprule
Layer & FoG-A & Rank & Layer & FoG-A & Rank \\
\midrule
\textbf{L0} & \textbf{1.1206} & 2 & L6 & 0.0917 & 4 \\
L1 & 0.0869 & 5 & L7 & 0.0552 & 7 \\
L2 & 0.0412 & 8 & L8 & 0.0365 & 10 \\
L3 & 0.0244 & 12 & \textbf{L9} & \textbf{2.4628} & 1 \\
L4 & 0.0779 & 6 & L10 & 0.0280 & 11 \\
\textbf{L5} & \textbf{0.0982} & 3 & L11 & 0.0389 & 9 \\
\bottomrule
\end{tabular*}
\end{table}

\begin{table}[t]
\centering
\caption{Where to align: Alignment layer with MuScriptor roll on SDD. L0/L2 are joint blocks; the remaining layers are fused blocks.}
\label{tab:layer_sweep}
\small
\setlength{\tabcolsep}{3pt}
\renewcommand{\arraystretch}{1.02}
\begin{tabular*}{\columnwidth}{@{\extracolsep{\fill}}lcc@{}}
\toprule
Layer & FAD $\downarrow$ & CLAP $\uparrow$ \\
\midrule
L0  & \textcolor{gray}{$0.5289 \pm 0.0032$} & $0.2431 \pm 0.0016$ \\
L2  & $\tabbest{0.4947 \pm 0.0037}$ & $\tabbest{0.2686 \pm 0.0015}$ \\
L4  & $0.5040 \pm 0.0030$ & $\underline{0.2603 \pm 0.0026}$ \\
L5  & $0.5202 \pm 0.0036$ & $0.2588 \pm 0.0014$ \\
L8  & $\underline{0.4980 \pm 0.0033}$ & $0.2587 \pm 0.0015$ \\
L9  & $0.5211 \pm 0.0045$ & $0.2457 \pm 0.0023$ \\
L11 & $0.5251 \pm 0.0036$ & $0.2459 \pm 0.0024$ \\
\midrule
Baseline & $0.5275 \pm 0.0021$ & $0.2220 \pm 0.0021$ \\
\bottomrule
\end{tabular*}
\end{table}
\begin{table*}[t]
\centering
\caption{ATTM objective comparison on 80 in-domain prompts. Borda ranks are
recomputed over the 17 official systems plus Ours; only the six official
finalists and the FluxAudio-S baseline are shown. Training cost follows ATTM's peak-FP16-throughput normalization.}
\label{tab:attm_leaderboard}
\small
\setlength{\tabcolsep}{4pt}
\renewcommand{\arraystretch}{1.1}
\begin{tabular*}{\textwidth}{@{\extracolsep{\fill}}lrrcccr@{}}
\toprule
\textbf{System} & \textbf{Params} & \shortstack{\textbf{A100-eq GPU-H}} & \textbf{FAD} $\downarrow$ & \textbf{CLAP} $\uparrow$ & \textbf{CCS} $\uparrow$ & \textbf{Rank} \\
\midrule
Submission \texttt{e07} & 402M & 148.9 & $0.417$ & $0.261$ & $0.867$ & $2$ \\
Submission \texttt{e01} & 189M & 25.1 & $0.577$ & $0.338$ & $0.863$ & $3$ \\
Submission \texttt{e05} & 499M & 20.0 & $0.487$ & $0.305$ & $0.800$ & $4$ \\
Submission \texttt{e08} & 450M & 190.3 & $0.495$ & $0.295$ & $0.804$ & $4$ \\
Submission \texttt{p05} & 2.4B & 60.0 & $0.514$ & $0.306$ & $0.800$ & $6$ \\
Submission \texttt{p00} & 502M & 48.0 & $0.557$ & $0.311$ & $0.796$ & $7$ \\
FluxAudio-S & \textbf{122M} & 25.8 & $0.757$ & $0.088$ & $0.592$ & $18$ \\
\midrule
\textbf{Ours} (MuScriptor roll, L2) & \textbf{122M} & \textbf{14.2} & $0.497$ & $0.315$ & $0.829$ & $\textbf{1}$ \\
\bottomrule
\end{tabular*}
\par\vspace{3pt}
\end{table*}


\par\vspace{-0.3em}
\section{MIDI-REPA design study}
\label{sec:design}
\subsection{Design protocol}
\label{subsec:experiments}

Since our setting is instrumental, we exclude SDD captions that mention vocals, leaving 629 prompts from 475 tracks. We select the symbolic target, alignment layer, and timestep interval sequentially, fixing each choice for the later studies. We measure audio quality with Fr\'echet Audio Distance (FAD) and prompt adherence with CLAP similarity~\cite{kilgour2018fr, laionclap2023}. The baseline uses the training configuration of Sec.~\ref{subsub:training_config} without alignment ($\lambda=0$). To reduce seed variance, we generate with eight matched seed sets and report the mean with 95\% confidence intervals.

\par\vspace{-0.5em}
\subsection{What to align: symbolic target}
\label{subsec:what}

Following the symoblic target construction in Sec.~\ref{subsub:target_extraction}, we compare Basic Pitch roll, MuScriptor merged roll, and MuScriptor 7-instrument cube at L4 over the full trajectory. Comparing the two rolls tests how the fine-grained transcription accuracy of the AMT system affects representation alignment. Comparing MuScriptor roll and cube tests whether alignment benefits more from instrument-agnostic note content or instrument-stem symbolic structure. We use L4 as the initial alignment point, following prior work favoring early-layer alignment~\cite{yu2024repa}.

As shown in Table~\ref{tab:teacher_comparison}, MuScriptor outperforms Basic Pitch roll on both FAD and CLAP, while Basic Pitch even yields worse FAD than the no-alignment baseline. This suggests that fine-grained transcription accuracy is critical, as noisy symbolic supervision can even degrade generation quality.

Among the roll and cube representations, the merged roll achieves better FAD, whereas the cube obtains higher CLAP score. We hypothesize that the cube preserves richer instrument-stem semantics, but is also more sensitive to noisy stem assignments. Given the above observation that noisy symbolic supervision can degrade generation quality, we use the \textbf{MuScriptor merged roll} in subsequent experiments.

\subsection{Where to align: alignment block}
\label{subsec:where}

Evaluating alignment at every layer is impractical. To narrow down the candidates, we first follow AG-REPA's Forward-only Gate Ablation (FoG-A)~\cite{zhang2026agrepa}, which measures the effect of removing each block on the predicted velocity field. As shown in Table~\ref{tab:foga}, the highest-scoring blocks are L9, L0, and L5. To better capture the trend across depth, we additionally sample representative blocks between and beyond these candidates, yielding the candidate set $\{\mathrm{L0,L2,L4,L5,L8,L9,L11}\}$.

As shown in Table~\ref{tab:layer_sweep}, alignment improves CLAP at every tested block and improves FAD at all blocks except L0. The first and last blocks perform relatively poorly, while, excluding these endpoints, layers up to L4 generally outperform the deeper layers, consistent with prior REPA findings that favor early-layer alignment~\cite{yu2024repa}, with L8 as an exception in FAD. Interestingly, none of the top FoG-A blocks are good alignment locations in our setting. Although it identifies layers that strongly affect the predicted velocity field, our results suggest that such functional importance does not necessarily translate into suitability for symbolic representation alignment. L2 overall achieves the best FAD and CLAP. We carry the MuScriptor roll \textbf{aligned at L2} forward to the flow-interval study.

\par\vspace{-0.5em}
\subsection{When to align: timestep interval}
\label{subsec:when}

We next ask \emph{when} symbolic alignment is most useful along the
generation trajectory. Does it primarily help establish
musical content before the melody is formed, refine and stabilize the
content after it has emerged, or remain useful throughout generation?

To locate this transition, we inspect the baseline generation trajectory
using the same 25-step Euler sampler described
in Sec.~2.2.3. Generation proceeds normally from $t=1$ to a selected
timestep $t$. From that state onward, however, we freeze the current guided
velocity and extrapolate directly to the clean endpoint:
\begin{equation}
    \hat{\mathbf z}_0(t)
    =
    \mathbf z_t
    -
    t\,\mathbf v_{\mathrm{CFG}}(\mathbf z_t,t,\mathbf c).
\end{equation}
Decoding $\hat{\mathbf z}_0(t)$ reveals the musical content implied at
timestep $t$, not the residual noise in the intermediate state
$\mathbf z_t$.\footnote{Directly decoding $\mathbf z_t$ indicates when residual noise has diminished enough for the sample to become perceptually music-like.}
\begin{figure}[t]
\centering
\includegraphics[width=0.95\columnwidth]{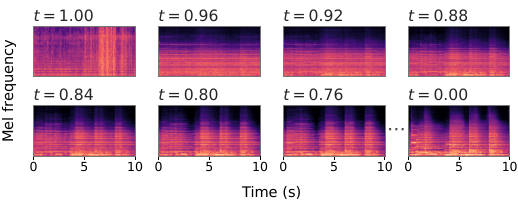}

\par\vspace{-1em}
\caption{
Extrapolated final outputs from different timesteps along the baseline generation trajectory for the prompt ``Atmospheric music featuring a soothing vibraphone, enhanced with gentle piano and ambient synth pads, evokes a calm, reflective mood with soft, resonant tones and a spacious, dreamy texture.'' Musical structure becomes largely established around $t=0.88$--$0.80$.
}
\par\vspace{-0.9 em}
\label{fig:timestep}
\end{figure}

As illustrated in Fig.~\ref{fig:timestep}, the predicted time--frequency
structure changes substantially at first but becomes largely established
around $t=0.88$--$0.80$. We therefore compare full-range alignment,
alignment before this transition $[0.8,1]$, after it $[0,0.8]$, and only
during the final refinement stage $[0,0.2]$. We additionally test early
boundaries at $0.88$ and $0.84$ to examine whether
continuing alignment beyond the transition is beneficial.

As shown in Table~\ref{tab:timestep_sweep}, aligning only before the melody is formed, $[0.8,1]$, or only after it has emerged, $[0,0.8]$, yields substantially worse FAD than full-range alignment, without improving CLAP. Restricting alignment further to only the final refinement stage, $[0,0.2]$, degrades both metrics substantially. Although the narrower early interval $[0.88,1]$ achieves the highest CLAP, its FAD is worse than the baseline, indicating a clear fidelity trade-off. Overall, \textbf{full-range alignment} achieves the best FAD while retaining a strong CLAP gain, suggesting that symbolic alignment is most effective when maintained throughout both melody formation and subsequent refinement. We therefore use $t\in[0,1]$ in the final configuration.
\par\vspace{-0.3em}
\section{Evaluation}
\label{sec:evaluation}
\par\vspace{-0.3em}
We evaluate the selected configuration on the ATTM's 80 in-domain prompts using the official objective metrics: FAD, CLAP similarity, and Concept Coverage Score (CCS)~\cite{hsieh2026academic}. Following the ATTM protocol, FAD is computed with LAION-CLAP music audio embeddings, while CLAP similarity uses audio--text embeddings from its music--speech--AudioSet model. Table~\ref{tab:attm_leaderboard} compares our model with the six finalists and the organizer's FluxAudio-S baseline provided by the organizer. \textbf{Our model remains competitive with substantially larger systems}, ranking fourth in FAD, second in CLAP, and third in CCS among the finalists shown. When inserted into the full 18-system objective comparison, the Borda rank recomputed jointly over FAD, CLAP, and CCS places our model at the top.
\begin{table}[t]
\centering
\caption{When to align: timestep interval with MuScriptor roll at L2 on SDD. Generation proceeds from $t=1$ (noise) to $t=0$ (music).}
\label{tab:timestep_sweep}
\small
\setlength{\tabcolsep}{3pt}
\renewcommand{\arraystretch}{1.02}
\begin{tabular*}{\columnwidth}{@{\extracolsep{\fill}}lcc@{}}
\toprule
Alignment interval $t$ & FAD $\downarrow$ & CLAP $\uparrow$ \\
\midrule
$t \in [0,1]$     & $\tabbest{0.4947 \pm 0.0037}$ & $\underline{0.2686 \pm 0.0015}$ \\
$t \in [0.88,1]$  & \textcolor{gray}{$0.5440 \pm 0.0042$} & $\tabbest{0.2858 \pm 0.0018}$ \\
$t \in [0.84,1]$  & $0.5201 \pm 0.0038$ & $0.2686 \pm 0.0036$ \\
$t \in [0.80,1]$  & $\underline{0.5200 \pm 0.0033}$ & $0.2641 \pm 0.0031$ \\
$t \in [0,0.8]$   & $0.5205 \pm 0.0029$ & $0.2683 \pm 0.0016$ \\
$t \in [0,0.2]$   & \textcolor{gray}{$0.5828 \pm 0.0035$} & $0.2581 \pm 0.0024$ \\
\midrule
Baseline           & $0.5275 \pm 0.0021$ & $0.2220 \pm 0.0021$ \\
\bottomrule
\end{tabular*}
\end{table}

Notably, our model uses only 122M parameters and 31.3 RTX 3090 GPU-hours of training (14.2 A100-equivalent GPU-hours based on peak FP16 Tensor throughput used in ATTM comparison), the smallest model and lowest training cost among the finalist shown. Moreover, the official 1,000-track FAD reference set was randomly drawn from the Jamendo training pool and hidden from participants, so systems trained on the full challenge corpus necessarily contain the FAD references in their training data. In contrast, we explicitly exclude all 1,000 reference tracks from training.


\par\vspace{-0.3em}
\section{Limitations and future work}
\par\vspace{-0.3em}
Our study is limited to a low-cost 122M-parameter setting and short-form generation; scaling to larger models and full songs remains future work. Because MIDI-REPA is agnostic to the AMT teacher, it can directly benefit from better transcription systems, which may make richer targets such as the instrument cube more effective. Beyond representation alignment, symbolic information could also serve as an output or an inference-time condition, pointing toward joint audio--MIDI generation and editable, controllable text-to-music models.

\par\vspace{-1.0em}
\section{Conclusion}
\par\vspace{-0.6em}
Modern text-to-music models learn musical structure largely through audio-space objectives alone. We show that symbolic supervision offers a simple, inexpensive, and complementary learning signal. Through a what--where--when design study, we find that aligning a MuScriptor piano roll at an early DiT layer over the full flow trajectory improves both audio quality and text adherence, suggesting that explicit musical structure is a practical training signal for text-to-music generation.

\clearpage
\section{Funding Acknowledgement}
\par\vspace{-0.3em}
This work is supported by grants from the National Science and Technology Council of Taiwan (NSTC 114-2628-E-002-013-MY3) and the Ministry of Education (MOE) of Taiwan (for Taiwan Centers of Excellence in Artificial Intelligence).

\section{Compliance with Ethical Standards}
\par\vspace{-0.3em}
This study did not involve experiments with human participants; therefore, ethical approval was not required.
\bibliographystyle{IEEEbib}
\bibliography{refs}
\end{document}